\documentclass[aps,prl,twocolumn,showpacs,floatfix]{revtex4}
\usepackage{epsf,psfig,epsfig,graphics}
\begin{document}

\title{Compact phases of polymers with hydrogen bonding}

\author{Antonio Trovato}
\affiliation{INFM - Dipartimento di Fisica `G. Galilei',
Universit\`a di Padova, Via Marzolo 8, 35131 Padova, Italy}
\author{Jesper Ferkinghoff-Borg}
\affiliation{EMBL, Structural Biology and Biocomputing
Meyerhofstrasse 1, 69117 Heidelberg, Germany}
\author{Mogens H. Jensen}
\affiliation{Niels Bohr Institutet, Blegdamsvej 17,
2100 K{\o}benhavn {\O}, Denmark}

\begin{abstract}
We propose an off-lattice model for a self-avoiding homopolymer
chain with two different competing attractive interactions, mimicking
the hydrophobic effect and the hydrogen bond formation respectively.
By means of Monte Carlo
simulations, we are able to trace out the complete phase diagram for
different values of the relative strength of the two competing interactions.
For strong enough hydrogen bonding, the ground state is a helical conformation,
whereas with decreasing hydrogen bonding strength, helices get eventually
destabilized at low temperature in favor of more compact conformations
resembling $\beta$-sheets appearing in native structures of proteins.
For weaker hydrogen bonding helices are not thermodynamically relevant
anymore. 
\end{abstract}


\maketitle

The collapse of a self-avoiding flexible polymer chain in a ``bad'' solvent has been
studied for many years \cite{Gro}. Following de Gennes seminal work on showing
the intimate connection of polymer collapse with tricritical
systems \cite{DesG}, most of the theoretical effort has been concerned
with the universal features of the $\theta$-point, the thermodynamic
second order transition between the swollen and the compact phase
\cite{vander}, this last phase being usually regarded as a structureless
liquid globule phase \cite{Gro}.
The possibility of more complex behavior in the compact phase has been
investigated only recently, revealing the existence, at lower
temperatures than the collapse gas-to-liquid transition, of a
liquid-to-solid and a solid-to-solid transition \cite{Zhou}.

On the other hand, protein molecules undergo similar transitions
between denatured, molten globule,
and native states, which are solid-like structures with a well defined
three-dimensional conformation \cite{Fersht}.
The main driving force of protein collapse is believed to be the
hydrophobic effect, which shields most of the non-polar side chains
in the core of the native protein structure from water \cite{Dill}.
This could indeed be grossly described as a ``bad'' solvent effect.
Yet, native structures of proteins are very peculiar, when compared
to typical compact conformations of self-avoiding polymer chains.
The benchmark of protein nativeness is perhaps the ubiquitous presence of
highly ordered local motifs, called secondary structures, known to be
stabilized by hydrogen bonding \cite{Pauling}.

In this Letter, we propose a minimal off-lattice 
homopolymer model, where a usual two-body isotropic attractive
interaction -mimicking the hydrophobic effect- is competing with a
directed attractive interaction mimicking the angular
dependence of hydrogen bonding \cite{Cre}. We consider a chain
of $N$ beads at positions $\vec{r}_i$ in the three-dimensional continuum
space $\mathbf{R^3}$, with $1\leq i\leq N$. The chain constraint is
enforced strictly, by keeping the distance
between consecutive beads along the chain constant and unitary,
$\left|\vec{r}_i-\vec{r}_{i-1}\right|=1$, for $2\leq i\leq N$,
while no other constraint is considered.
We model the hydrophobic effect \cite{water} by
considering a pair-wise attractive square well potential with a hard wall, $E_{hp}
= \sum_{0\leq i-1<j\leq N} V_{hp}\left(\left|\vec{r}_i-\vec{r}_j\right|\right)$,
with:

\begin{equation}
V_{hp}\left(r\right)=\left\{\begin{array}{lll} \infty &
\qquad {\rm for} \quad r\leq\sigma  \\  -1 & \qquad
{\rm for} \quad \sigma\leq r \leq \lambda \\ 0 & \qquad
{\rm for} \quad r \geq \lambda \end{array}\right. \; ,
\label{Nonpolen}
\end{equation}
where $\sigma$ is the hard-core radius of each bead, and $\lambda$ is the
range of the attractive interaction. In the following we will always
consider the case $\sigma=1$, $\lambda=1.5$, as in \cite{Zhou}.

In order to model hydrogen bonding, we need to break isotropy and favor
a preferred direction between the two `hydrogen-bonded' beads.
We use the same type of directed interaction proposed by
Chen and Kemp \cite{Jeff}, so that the two planes, each containing one of
the two hydrogen-bonded beads and its nearest neighbors along the chain,
will both be preferibly orthogonal to the contact vector between them \cite{Multibody}:
$E_{hb} = \sum_{2\leq i<j+1\leq N}
V_{hb}\left(\vec{r}_i-\vec{r}_j,\vec{u}_i,\vec{u}_j\right)$, where
$\vec{u}_i=\left(\vec{r}_{i+1}-\vec{r}_i\right)\times
\left(\vec{r}_i-\vec{r}_{i-1}\right)$, and:

\begin{equation}
V_{hb}\left(\vec{r},\vec{u}_i,\vec{u}_j\right) =
0.5 \left( \left|\hat{r}\cdot\hat{u}_i\right|^m +
\left|\hat{r}\cdot\hat{u}_j\right|^m\right)
V_{hp}\left(\left|\vec{r}\right|\right) \;, 
\label{Hbonden}
\end{equation}
where $\hat{\cdot}$ denotes normalized vectors.
The directionality degree of hydrogen bonding is controlled by the exponent
$m$; a large value corresponds to a strong `directionality'. 
We have mainly studied the $m=12$ case, since lower values
of $m$ do not favor protein-like secondary structures in this
parametrization.

The interplay between hydrophobic collapse and hydrogen bonding is
controlled by the relative strength $\alpha$ between the two interactions
when the following total Hamiltonian is considered:

\begin{equation}
{\cal H}_{\alpha} = E_{hp} + \alpha E_{hb} \; .
\label{toten}
\end{equation}

Whenever two beads come into contact, i.e. their mutual distance
falls within the well, they always gain a negative unitary energy
contribution from $E_{hp}$. A further negative contribution could come
from $E_{hb}$, depending on how well the hydrogen bond is formed between
them, ranging from $0$, in the worst case, to $-\alpha$, in the best one.
Thus, from the microscopic point of view the two energy terms are cooperative.
If hydrogen bonding is switched off, $\alpha=0$, we are back to the
usual case of isotropic pairwise attraction considered in \cite{Zhou},
which yields a compact groundstate with no secondary structures.
In the other limit of no hydrophobic interaction, $\alpha=\infty$,
the ground-state has already been shown to be a long straight helix, 
when $m\geq6$ \cite{Jeff}. Since the ground state differs significantly
in the two limiting cases, one should actually expect a
non-trivial competition between the two energy terms for 
intermediate values of $\alpha$, despite the microscopic cooperativity.
This competition is induced at a global macroscopic level as a
consequence of chain connectivity and excluded volume constraints \cite{Glgl}.
In this Letter, we will focus on its thermodynamic implications.

Our results qualitatively agree with previous work on an
analogous lattice model, where hydrogen bonding was mimicked via the
introduction of rotating spins \cite{BMST}.
We remark, nonetheless, that the extension of such results to our off-lattice
model is higly non trivial, since the geometrical order implicit in the
lattice structure could mask or enhance artificially secondary structure
formation. As an example, while isotropic compaction of a homopolymer chain
on a cubic lattice is sufficient to produce some amount of secondary
structure \cite{Chan}, this is not true for an off-lattice homopolymer
\cite{Onu,Greg}.

Our aim is to determine, by means of Monte Carlo simulations,
the density of states $\rho\left(E\right)$ of a polymer chain
with the Hamiltonian (\ref{toten}), so that the partition function 
$Z_N\left(T\right)$ of a $N$-bead chain at reduced temperature $T$
can be easily reconstructed: $Z_N\left(T\right)=\sum_E\rho
\left(E\right)\exp\left[-E\!/T\right]$.
We have employed a set of standard moves currently used in simulations
of polymer chain; pivot, crankshaft, and reptation moves \cite{Sokal}.
In order to avoid trapping in local energy minima,
we have employed a novel simulation method,
based on generalized ensemble techniques \cite{Jesper}. The key notion,
using generalized ensembles, is that
a proper reweighting of temperature as a function of energy should allow
the chain to escape from such energy minima \cite{Mult}.
The method lends itself in a natural way to be formulated within an
iterative convergence scheme, and the possibility of properly
employing the statistical information from more different steps of such
scheme greatly increases its effectiveness \cite{Jesper}.

Nevertheless, the presence of frustration provides an inherent
limitation to such method, since it is based on the knowledge of local
properties of the phase space, and the competition between different
energy terms results in different regions of the phase space sharing
the same total energy but having different local densities of states.
This turns out to be the case within our model, causing a very slow
convergence to equilibrium. Therefore,
we have introduced a finer, two-dimensional representation of the full
multi-dimensional phase-space, by identifying a conformation
through both its hydrophobic energy $E_{hp}$ and hydrogen bond energy
$E_{hb}$. Our simulation method can be
easily adapted in order to compute the density of states $\rho\left(E_{hp},
E_{hb}\right)$ as a function of both energy terms. Details on the employed
reweighting scheme will be published elsewhere \cite{long}.
In this way, not only convergence to equilibrium is more easily obtained,
but the partition function
$Z_N\left(T,\alpha\right)=\sum_{E_{hp},E_{hb}}\rho\!\left(E_{hp},E_{hb}\right)
\exp\left[-\!\left(E_{hp}+\alpha E_{hb}\right)\!/T\right]$, and hence
any other relevant thermodynamic quantity, can now be reconstructed for
any given value of the relative strength $\alpha$ between the two competing
energy terms. The effectiveness of this sampling strategy shows that
ther two energy terms serve as relevant order parameters.

\begin{figure}
\centering
\includegraphics[width=3.4in]{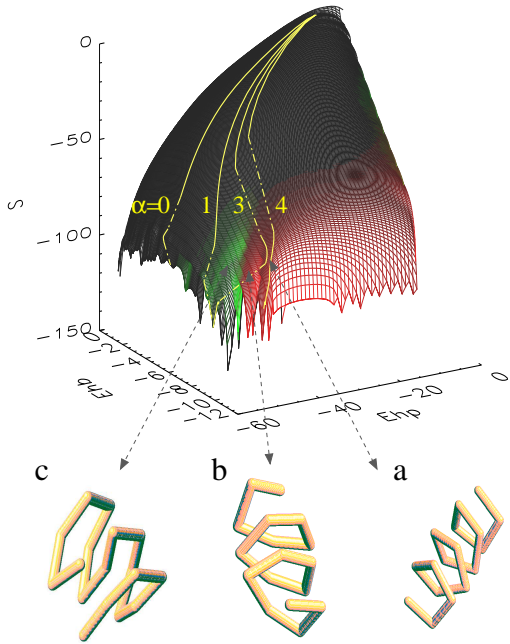}
\caption{Entropy surface plot $S\left(E_{hp},E_{hb}\right)$
in the ($E_{hp}$,$E_{hb}$) plane. Secondary structure content
order parameters are shown in color scale, red for helices $<\Sigma_h>$,
green for sheets $<\Sigma_{bs}>+<\Sigma_{as}>$. The lighter the color
the higher the order parameter. The yellow lines on the
entropy surface show the average hydrophobic and hydrogen bond energies
parametrized as a function of temperature for $\alpha=0,1,3,4$.
The dashed portions of the curves refer to a first order transition,
which is identified by looking at the free energy contor plot, as in
Fig. $3$, and simply connect the two competing free energy minima.
Typical conformations populating relevant entropic ridges are also shown.}

\end{figure}

The main result of this work, obtained for a chain with $N=17$ beads,
is shown in Fig. $1$. The logarithmic density of states
$S\left(E_{hp},E_{hb}\right)=\ln\left[\rho\left(E_{hp},E_{hb}\right)\right]$,
the microcanonical entropy, is represented as a surface plot in the
employed two-dimensional representation of the conformational space.
Effective free energy landscapes can easily be reconstructed
within the same representation:

\begin{equation}
F_{\alpha}\left(T,E_{hp},E_{hb}\right)=
\left(E_{hp}+\alpha E_{hb}\right)\!/T-S\left(E_{hp},E_{hb}\right),
\label{freen}
\end{equation}
where the free energy $F_{\alpha}\left(T,E_{hp},E_{hb}\right)$
is given in dimensionless units.
The surface plot in Fig. $1$ can thus be interpreted as the opposite
of the free energy landscape at infinite temperature, so that free
energy valleys are seen as entropic ridges.
In the phase space region with the lowest values of both energy terms,
the entropy surface exhibits a rich yet regular structure,
which is going to play a crucial role in determining the thermodynamic
properties at low temperature. Three different ridges, separated
by non-convex regions of the entropy surface corresponding to free energy barriers, can be identified.

The properties of the conformation ensembles populating such entropic ridges,
or free energy valleys, can be readily identified by computing several
order parameters, which measure the compactness degree and the amount of
secondary structure content.
Compactness is usually measured by means of the square gyration radius:

\begin{equation}
R_g^2 = \sum_{i=1}^N \left(\vec{r}_i-\vec{r}_{cm}\right)^2 /N \; ,
\label{ordpgr}
\end{equation}
where $\vec{r}_{cm}=\sum_{i=1}^N\vec{r}_i/N$ is the center of mass
vector.
As for secondary structures, we define the helical content
of a conformation as:

\begin{equation}
\Sigma_h = \sum_{j-i=3}^5
\left[\left(V_{i-1,j-1}+V_{i,j}+V_{i+1,j+1}\right)/3\right]^m \; ,
\label{ordhel}
\end{equation}
and the parallel and antiparallel sheet content similarly:

\begin{eqnarray}
\Sigma_{ps} & = & \sum_{j-i\geq6}
\left[\left(V_{i-1,j-1}+V_{i,j}+V_{i+1,j+1}\right)/3\right]^m \; ,
\label{ordbetapar} \\
\Sigma_{as} & = & \sum_{j-i\geq5}
\left[\left(V_{i-1,j+1}+V_{i,j}+V_{i+1,j-1}\right)/3\right]^m \; ,
\label{ordbetaapar}
\end{eqnarray}
where $V_{i,j}=V_{hb}\left(\vec{r}_i-\vec{r}_j,\vec{u}_i,\vec{u}_j\right)$,
($0\leq V_{i,j}\leq1$) measures to what extent a hydrogen bond is
formed between beads $i$ and $j$. Each term in the above sums
(\ref{ordhel}),(\ref{ordbetapar}),(\ref{ordbetaapar}), again
between $0$ and $1$, measures to what extent the $\left(i,j\right)$ pair can
be considered the center of a local portion of a given secondary structure.
Within our definitions a simple hydrophobic contact, which is not a
good hydrogen bond, does not contribute to secondary structure counting.

As is shown in Fig. $1$, the ridges in the entropy surface are
associated, with increasing number of hydrophobic contacts and
decreasing number of hydrogen bonds, to helices with $4$ beads per turn,
to helices with $5$ beads per turn, and to sheet-like conformations,
respectively. The gyration radius decreases accordingly \cite{long},
since long straight helices are extended objects.
Whereas helices (a) and (b) are indeed representative of the two
helical ridges, the sheet-like conformation (c) is just one among many
possible different representatives.
In this region we expect the occurrance of many different free
energy minima, possibly giving rise to glassy behavior.
Such frustration is of course not resolved within our bivariate
parametrization.

\begin{figure}
\centering
\includegraphics[width=3.4in]{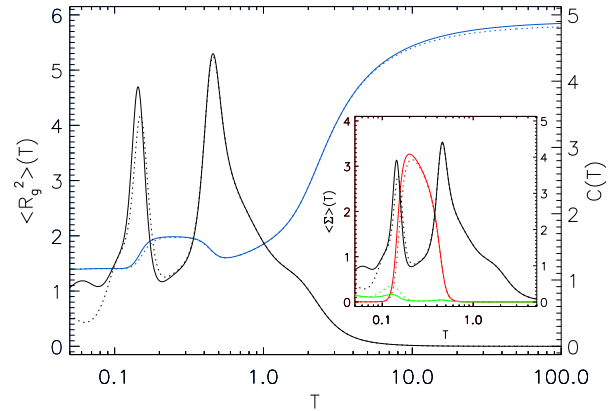}
\caption{Specific heat per monomer $C$, black line, and mean square
gyration radius $<R^2_g>$, blue line, as a function of reduced temperature
$T$ in logarithmic scale, in the $\alpha=3$ case.
In the inset, the specific heat per monomer, black line,
is again shown together with the secondary structure order parameters
$<\Sigma_h>$, red line, and $<\Sigma_{bs}>+<\Sigma_{as}>$, green line.
Dotted lines show the same quantities as computed from a second independent
simulation.}
\end{figure}

We now discuss in detail the case $\alpha=3$.
The mean square gyration radius $<R_g^2>$, and the specific heat per monomer
$C = T^{-2}\left(<{\cal H}_{\alpha}^2> - <{\cal H}_{\alpha}>^2\right)/N$
are shown in Fig. $2$. The behavior of the secondary
structure order parameters, $<\Sigma_h>$ and $<\Sigma_{ps}>+<\Sigma_{as}>$,
is shown in the inset. The curves resulting from two
different independent simulations are shown; the accuracy is quite good
down to temperatures as low as 0.1. The specific heat curve exhibits, with
decreasing temperature, one shoulder and two higher and sharper peaks.
Specific heat peaks are usually related to a
phase transition, but care should of course be taken in generalizing
results from such a small system (see e.g. \cite{Grass} for a detailed
discussion of problems arising in finite-size scaling of $\theta$-collapse).
For a finite-size analysis of such transitions, we show the free energy
contour plots at the corresponding temperatures in Fig. $3$.

As signalled by the decrease of the gyration radius, the specific heat shoulder is
related to the collapse of the chain from the swollen high temperature phase.
The first peak close to the shoulder corresponds to a sharp increase of the
gyration radius, and is related to the formation of many hydrogen
bonds and to the appearance of helical structure, whereas the hydrophobic
energy is almost not changed.
The free energy contour plot clearly shows the existence of two
competing minima, so this globule-to-helix transition is first order.
The second peak is marked by a sharp decrease in the gyration radius,
and is related to the breaking of helices and to the formation of
sheet-like structures, as (c) in Fig. $1$.
Sheet-like conformation are compact objects, having less hydrogen bonds
but more hydrophobic contacts than helices.
This last transition is again first order, as shown in the
free energy contour plot. {\it Three} different minima
are present, originating from the entropic ridges identified in Fig. $1$,
but helices with $5$ beads per turn (b) never get efficiently populated,
since they suffer competition from either side.

The very structure of the specific heat, one shoulder
and then two peaks with decreasing temperature, is similar to what is
found in the usual $\alpha=0$ case, even if the thermodynamically
stable phases are completely different.
It is tempting to interpret our results within the same overall framework
proposed in \cite{Zhou}, that is to say with decreasing temperature the
chain first undergoes a gas-to-liquid collapse, then a first order
liquid-to-solid transition, and finally a solid-to-solid transition
which is again first order (in the absence of hydrogen bonding, 
the last transition is a continuous polymorphic transition \cite{Zhou}).
In our model, the possibility of hydrogen bonding simply acts in selecting
helices and sheets among all possible solid crystalline conformations.

\begin{figure}
\centering
\includegraphics[width=3.4in]{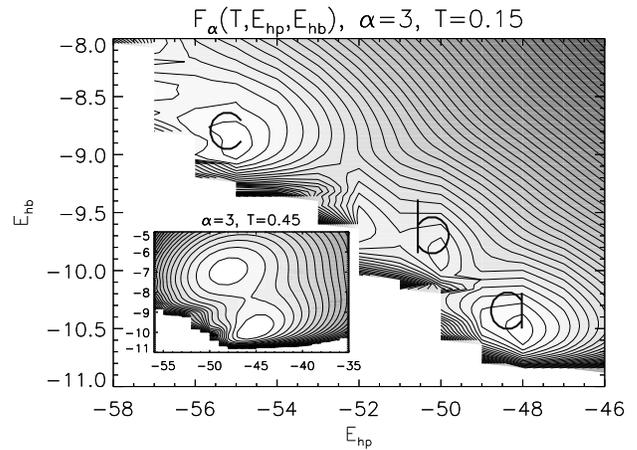}
\caption{Contour plots at different temperatures, in the $\alpha=3$ case,
of the effective free energy $F_{\alpha}\left(T,E_{hp},E_{hb}\right)$,
eq. (\ref{freen}), in the ($E_{hp}$,$E_{hb}$) plane.
The temperatures $T=0.46$, $T=0.14$, correspond
to the specific heat peaks seen in Fig. 2.
The spacing between consecutive levels in each contour plot is unitary,
and corresponds to a difference of $k_BT$ in physical units.
The darker the color, the higher the free energy value.
Letters refer to entropic ridges and conformations in Fig. 1.}
\end{figure}

In Fig. $1$ we have also summarized the different thermodynamic {\em static}
properties of the polymer chain when $\alpha$ is varied. 
The yellow lines can be thought of as dynamical trajectories only in the
{\em infinitely slow} cooling case. Actual dynamics
does {\em not} take place within the effective free energy landscape
(\ref{freen}), since kinetic barriers in the full phase space
are smoothed over by the coarse-graining of our representation.
This is most likely the case for the helix-to-sheet transition, 
where we expect the underlying energy landscape to be much more roughed.

All trajectories in Fig. $1$ start from a common point at infinite temperature,
but then explore different regions of the phase space, according to
different strengths of hydrogen bonding. Nevertheless a
collapse transition, related to the shoulder in the specific heat, is
common to all $\alpha$ values, and moreover takes place in all cases
for similar values of the competing energies.
At lower temperatures, hydrogen bond strength greatly affects the 
thermodynamic behavior. When $\alpha=0$, the conformational ensemble
populated at low temperature does not include, as expected, the regions
of the phase space with a high content of secondary structure.
As in \cite{Zhou}, two further transitions are present,
liquid-to-solid and solid-to-solid. If $\alpha=1$, after the last
solid-to-solid transition the sheet-like region of the phase space becomes
efficiently sampled at low temperatures. If $\alpha=3$, as we have already
seen, the chain first undergoes a transition from the liquid globule phase
to the helical region and then to the sheet-like region. Both transitions
are first order.
Finally, if $\alpha=4$, hydrogen bonds are strong enough to produce a helical
ground state, as in the $\alpha=\infty$ case \cite{Jeff}.

Note that only the marginal border of the `green' sheet-like region is
thermodynamically relevant at low temperatures
(see also the small increase of the order parameter in Fig. $2$).
It is believed that $\alpha$-helices are more likely to be
formed by residues with small side chain groups,
whereas the loss in conformational entropy suffered
by bigger side chain groups, when arrenged in helical conformation,
favors the formation of $\beta$-sheets \cite{Rose}.
This general picture is consistent with our results. In fact,
no side groups are present and helices are indeed
entropically favored, since they sit on the top of a ridge in the
entropy surface, whereas sheet-like conformations do not.

To summarize, we have introduced a simple model for an off-lattice
self-avoiding
polymer chain with two competing attractive inteactions, isotropic and 
directionalized. By means of Monte Carlo simulations we have determined
the density of states of the chain within a two-dimensional representation
of the phase space, and hence the phase diagram for different values of
the relative strength of the two competing energies.
If the directionalized interaction is strong enough, different
conformational ensembles compete closely with each other at low temperature,
which have peculiar proteinlike features, such as helices and sheets.

It is a pleasure to acknowledge enlightening discussions with
A.~Maritan, C.~Rischel, K.~Sneppen and G.~Tiana.

\eject

\begin{thebibliography}{99}
\bibitem{water}
A proper tracing out of the solvent molecules degrees of freedom would
indeed yield a temperature dependent pair-wise potential. We neglect such
temperature dependence for the sake of simplicity, as is usually done
in similar studies \cite{Zhou}.
\bibitem{Multibody}
Note, that isotropy breaking involved in hydrogen bond formation
introduces effective multibody correlations.
\end{thebibliography}
\end{document}